\newcommand{\zi}{\mbox{$z _{1}$}}

\documentstyle[12pt]{article}

\newcommand{\be}{\begin{equation}}
\newcommand{\br}{\begin{eqnarray}}
\newcommand{\ee}{\end{equation}}
\newcommand{\er}{\end{eqnarray}}

\newcommand{\p}{\mbox {$ \partial$}}

\begin{document}
\title{The Proper Time Equation and the Zamolodchikov Metric}
\author{B. Sathiapalan\\ {\em
Physics Department}\\{\em Penn State University}\\{\em 120
Ridge View Drive}\\{\em Dunmore, PA 18512}}
\maketitle
\begin{abstract}
The connection between the proper time equation and the Zamolodchikov
metric is discussed. The connection is two-fold:
First, as already known, the proper time
equation is the product of the Zamolodchikov metric and the renormalization
group beta function.  Second, the condition that
the two-point function is the Zamolodchikov
metric, implies the proper time equation.
We study the massless vector of the open string in detail.
In the exactly calculable case of a uniform electromgnetic
field strength we recover the Born-Infeld equation.
We describe the systematics of the perturbative evaluation
of the gauge invariant proper time equation for the massless vector
field. The method is valid for non-uniform
fields and gives results that are exact to all orders in derivatives.
As a non trivial check, we show that in the limit of uniform
fields it reproduces the lowest order Born-Infeld equation.
\end{abstract}
\newpage
\section{Introduction}

        The sigma model renormalization group approach to computations
in string theory \cite{l,cc,as,ft}, and its various generalizations,
[5-25], have been very fruitful.  In its original version it involves
calculating various $\beta$-functions for the generalized coupling
constants of the sigma model.  The equations of motion that one obtains
from an S-matrix calculation are proportional to the $\beta$-function,
the proportionality factor being the Zamolodchikov metric\cite{zam}.
The fact
that there must be a proportionality factor, was inferred in \cite{ac}
by calculating the $\beta$-function exactly for a constant electromagnetic
field background, and showing that it could not be obtained from an action
unless multiplied by a prefactor.  Polyakov outlined an argument that
showed that the proportionality factor is the Zamolodchikov metric
\cite{polya}. In \cite{bspt} this was demonstrated in detail
for the tachyon.

In \cite{bspt} it was also shown that an equation, called the proper
time equation (which is similar to the proper time equation for a point
particle\cite{rf,js,yn,pm,sm,ct})
is in many ways easier to calculate than the $\beta$-function,
especially for the tachyon and massive modes.  The calculation
of the various terms of the proper time equation
is very similar to an S-matrix
calculation in first quantized string theory.
It is therefore quite easy to see that it gives the full equation of
motion of the string modes (i.e. including the Zamolodchikov metric
prefactor)\cite{bspt}.  In \cite{bspt} the discussion was confined to
the tachyon.  More recently \cite{bsrg},
it was applied to the massless vector
field where some results are known \cite{ac,ft1,t,do}.  In
particular it is known that in the limit of constant field strength,
the equation of motion of the photon is that derived from the
Born-Infeld action,
\be   \label{1.1}
\frac{\delta}{\delta A_{\mu}}\int d^{D}X \sqrt{Det(I+F)}
=\sqrt{Det(I+F)}(I-F^{2})^{-1}_{\mu \nu}\p _{\rho} F_{\nu \lambda}
(I-F^{2})^{-1}_{\lambda \rho} =0
\ee
The exact $\beta$-function for this theory in this limit is also known
\cite{ac}
\be         \label{beta}
\beta _{\nu} = \p _{\rho} F_{\nu \lambda} (I-F^{2})^{-1}_{\lambda \rho}
\ee
The prefactor in (\ref{1.1}) is thus the Zamolodchikov metric.

In \cite{bsrg} leading corrections to Maxwell's equation were calculated
using the proper time formalism. It was shown that the zero momentum
limit of these corrections agreed with the Born-Infeld result - namely
that they come from the variation of
\be     \label{1.5}
Tr F^{4} -1/4 (TrF^{2})^{2}
\ee
Actually this was demonstrated in a gauge fixed calculation
involving transverse, on-shell photons satisfying $k^{2}=k.A=0$.
The gauge invariant calculation is a little more tricky and only
partial results were presented - it was shown that the $(Tr F^{2})^{2}$
is present with a non-zero coefficient.  This term is significant
because it does not show up in the $\beta$-function calculation \cite{ac}.
(See equation  \ref{beta}).  The reason for this is explained in \cite{bsrg}.
The fact that it shows up in the proper time equation is evidence
that it is the complete equation.

It should also be pointed out that the proper time equation gives
results that are exact to all orders in derivatives.  This is unlike
sigma model perturbation theory where one performs a derivative
expansion.  In particular the equation is valid for finite
values of momenta and all the massive poles are manifest just
as in an S-matrix calculation. Corresponding calculations
of $\beta$-functions are much harder because one has to
disentangle various subdivergences \cite{bspt,bsrg}.
One look at the $\beta$-function
calculation for the Sine-Gordon theory \cite{amit} or the
tachyon \cite{ds} should convince anyone of this.

In this paper we would like to set up the systematics of evaluating
the gauge invariant proper time equation for the massless vector field.
Our perturbation series will be in the field strength.  At each order
in the field strength we will write down a formal expression
that is exact to all orders in the momentum of the vector field.
The formal expression is very
similar to the Koba-Nielsen representation.  In some regions we
can evaluate it exactly in terms of Gamma functions and related functions.
The results, by construction, agree with S-matrix when the fields
are exactly on shell.  The hope is that the proper time equation is valid
even when the fields are not exactly on-shell.  When one is far off-shell
one has to resort to other devices (perhaps as in \cite{bsfc})
or string field theory \cite{wsft}.

There are some simplifications and also some subtleties that one must
be aware of when actually doing the calculation.  In the Koba-Nielsen
representation the integration variables satisfy $0\leq z_{1}\leq z_{2}
\leq z_{3} \leq...\leq z_{N} \leq 1$.  We could set up the problem
in the same way for
the proper time equation also - and this was done in \cite{bspt}.
In $\beta$-function calculations, on the other hand, the range
of integration is from $- \infty $ to $+ \infty$.  There are
certain advantages to this: for instance the following identity for
2-D Green functions (easily verified by going to momentum space)
enormously simplifies the $\beta $-function calculation \cite{ac,ft1}
\be      \label{1.2}
\int _{-\infty} ^{\infty} dz^{\prime} \p_{z} G(z-z^{\prime})
\p _{z^{\prime}}G(z^{\prime}-z^{\prime \prime}) = \delta (z-z^{\prime \prime})
\ee
It is crucial here that the range of integration be from $-\infty$
to $+\infty$.  We will see that we can fruitfully modify the
range of integration to ($-\infty , +\infty$) in the proper time equation
also.  Another subtle issue is the zero momentum limit. It will
turn out that in some of the calculations it is necessary to keep
momenta non-zero initially and then take the limit of zero momentum.
This is because many of these expressions are divergent in the zero
momentum limit and there are cancellations between poles and zeros that
one needs to keep track of.

One can also study the zero momentum limit directly as in \cite{ac,ft1}.
In this case one can use the exact Geen functions and check that the
proper time equation gives the full equation.  One can also separately
calculate the Zamolodchikov metric and confirm earlier results.

This paper is organized as follows.
In Section 2 we describe in general terms the connection between
the proper time equation and the Zamolodchikov metric.  In Section 3
we look at the case of a uniform electromagnetic field and obtain
the full Born-Infeld equation.  In Section 4 we describe gauge invariant
perturbation theory for the proper time equation in the case of
non-uniform electromagnetic field.  In Section 5 we take the zero
momentum limit of the general result and show that there is agreement
with the results of Section 3.  We conclude in Section 6 with a summary
and some comments.

\newpage
\section{Proper Time Equation and the Zamolodchikov Metric}
\setcounter{equation}{0}
The proper time equation in its simplest form (for the tachyon) is
\be                  \label{pt}
\left. \int d^{D}k \{ \frac{d}{d\ln z} z^{2} <V_{p}(z) V_{k}(0)>\}
\right | _{\ln  z=0}
\Phi (k) =0 .
\ee
The evaluation of the derivative at $\ln z =0$ ensures that higher
powers of $\ln z$ do not contribute to the equation.
In (\ref{pt}) $\Phi$ is the tachyon field associated with the vertex
operator $V$.  We have indicated the momentum dependence of the
vertex operator by the subscripts.  The full tachyon perturbation
is
\be              \label{2.2}
\int dz d^{D}k e^{ik.X(z)}\Phi (k) \equiv \int dz d^{D}k V_{k}(z)
\Phi (k) .
\ee
We have to assume that $\Phi (k)$ is non zero only in a small region
around $k^{2}=2$ in order for (\ref{pt}) to make sense.  We are
assuming that the fields are almost on-shell , i.e. they are
marginal perturbations.  This is a serious limitation in all these
approaches.  To get around this one needs to retain a finite
cutoff on the world sheet \cite{bspt,bsos,bsfc}.  We will not
discuss this aspect of the problem in this paper and henceforth
we will assume that all fields are almost on-shell.  For massless
fields this means $k^{2}\approx 0$.  Note that this does not mean
that $k \approx 0$.  Thus $k$ can be finite (but smaller than
$\frac{1}{\sqrt \alpha '}$)\footnote{If $k\geq\frac{1}{\sqrt \alpha '}$
we run into issues about non-renormalizability of the model
and the inclusion of massive modes \cite{ds2}}.

The vacuum expectation value in (\ref{pt}) is evaluated with the
usual Polyakov measure but with (\ref{2.2}) added as a perturbation.
Thus one can evaluate (\ref{pt}) as a power series in $\Phi$.
Furthermore the range of integration is from 0 to 1.  It is also
important to regularize (\ref{pt}) with an ultraviolet cutoff.
One simple and convenient way is to alter the limits of integration
to ensure that there is a minimum spacing $a$ between two vertex
operators.  This has the effect of subtracting pole terms
which is exactly what one wants in an effective action \cite{bspt}.

It was shown in \cite{bspt} that the above prescription reproduces
the equation of motion.  The argument is very simple:  In an
S-matrix calculation, because of SL(2,R) invariance, one would
hold three of the vertex operators fixed - say at 0, $z$ and $z_{1}$
with $0  \leq z_{1}\leq z$. The result of integrating
$z_{2},z_{3},...,z_{N}$ is to produce the N-particle S-matrix
element multiplied by $z^{-1+\epsilon}(z-z_{1})^{-1+\delta}z_{1}
^{-1+\gamma}$, where $\epsilon , \delta , \gamma $ are infinitesimals
that vanish when the particles are exactly on-shell.  In an S-matrix
calculation this factor
would then exactly cancel the Jacobian from SL(2,R) gauge fixing:
$z(z-z_{1})z_{1}$. However in (\ref{pt}) there is a further integration
over $z_{1}$. On doing this integral and looking at the $\ln z$
deviation from $1/z^{2}$ and taking the limit $\epsilon = \delta
=\gamma =0$ we merely get a factor 2, which multiplies the S-matrix
element.  The effect of the ultraviolet regulator is as mentioned
before, to subtract pole terms from the S-matrix.  To show that
this is also equal to the Zamolodchikov metric times the
$\beta$-function requires a little more work.  We refer the reader
to \cite{bspt,polya}.

One can also generalize (\ref{pt}) by adding indices to the
vertex operators and fields. These could be Lorentz indices
or a general index indicating  the different modes of a string.
Thus we get the general proper time equation:
\be       \label{2.3}
\left. \sum _{J}  \{ \frac{d}{d\ln (z/a)} z^{2} <V_{I}(z) V_{J}(0)>\}
\right| _{\ln (z/a)=0}
\Phi ^{J} =0 .
\ee
This is a set of equations, one for each value of the index $I$.
Note that, with an ultraviolet regulator in the form of a short
distance cutoff, $a$, $\ln z$ is actually $\ln (z/a)$.  Thus we are
evaluating the derivative with respect to $\ln z$ at $z=a$.
(This point was pursued further in
\cite{bspt,bsos,bsfc} to study off-shell theories.)  If one
works with renormalized quantities this could be rewritten
as $z=b$ where $b$ is some renormalization scale.

The two-point function is related to the Zamolodchikov metric.
Consider the following general expression for the two-point
function of two VO's, both marginal (i.e. dimension 1):
\be      \label{2.4}
<V_{I}(z) V_{J}(0)> = \frac{G_{IJ}}{z^{2}} +  \frac{H_{IJ}}{z^{2}} \ln (z/a)
+ O(\ln ^{2}(z/a))
\ee
In general $G_{IJ}(\phi)$ and $H_{IJ}(\phi)$ are functions of the background
fields - the coupling constants of the two dimensional theory.
Here $G_{IJ}$ is the Zamolodchikov metric.  Thus
\be         \label{2.5}
<V_{I}(a) V_{J}(0)> = \frac{G_{IJ}}{a^{2}}
\ee
Comparing (\ref{2.4}) with (\ref{2.3}) the proper time equation is
\be        \label{2.6}
\sum _{J} H_{IJ}\Phi ^{J} =0
\ee
Thus when the two point function is equal to the Zamolodchikov
metric (upto terms of $O(\ln (z/a)^{2})$) the proper time equation
is satisfied.  The reverse is not quite true since there is a sum
over the index $J$.  We can however say that
the two point function and the Zamolodchikov metric
viewed as  matrices, are equal when acting on the subspace of
solutions of the proper time equation. Thus acting on this subspace:
\be        \label{2.7}
z^{2}<V_{I}(z) V_{J}(0)> = G_{IJ} + O(\ln ^{2}(z/a))
\ee
The original motivation for the proper time equation was quite different.
But we find this connection with the Zamolodchikov metric
very interesting.

We can go a little further. We also know that
\be          \label{2.8}
H_{IJ} \Phi ^{J} = G_{IJ} \beta ^{J}
\ee
This is just the statement made earlier that the proper time equation is
the Zamolodchikov metric times the $\beta$-function.  Here
\be         \label{2.9}
\beta ^{J} \equiv - \frac{d \Phi ^{J}}{d \ln a}
\ee
 Thus we find
\be         \label{2.10}
z^{2} <V_{I} (z) V_{J} (0)> \Phi ^{J} = G_{IJ} \Phi ^{J} +
G_{IJ} \beta ^{J} \ln (z/a)
\ee
If we let $z=\lambda a$, where $\lambda = 1+ \epsilon \: \: ,
\epsilon \approx 0$ then the RHS of (\ref{2.10}) can be written as
\be        \label{2.11}
G_{IJ} \Phi ^{J} + G_{IJ} \delta \Phi ^{J}
\ee
where

\be        \label{2.12}
\delta \Phi ^{J} = \beta ^{J} \ln \lambda
\ee
which is the change in $\Phi ^{J}$ under the scale change $a \rightarrow
\frac{a}{\lambda}$.
Thus
\be
z^{2} <V_{I}(z) V_{J} (0) > \Phi ^{J} = G_{IJ} (\Phi^{J} + \delta \Phi ^{J}).
\ee
which is nothing but the statement that the evolution of the string
is really a renormalization group flow in the two dimensional theory.

The connection with the Zamolodchikov metric also suggests
a connection with the background independent formalism of
\cite{w,wl}.  In fact one could try to transcribe the proper time
equation into the BRST formalism.
In the BRST formalism one would have to include the ghosts.  The
renormalization group transformation would then be replaced by a
BRST transformation.  This should give us an equation similar
to that derived in \cite{w,wl,s}.

This concludes our review and our discussion of the proper time
equation in general terms. In subsequent sections we specialize
to the case of the massless vector field in the open string.

\newpage
\section{Uniform Electromagnetic Field}
\setcounter{equation}{0}

We now turn to the massless vector field in the limiting case
of (an almost) uniform field strength.  This has been discussed
in \cite{ac,ft1,t}. The vector field perturbation added to the
Polyakov action is
\be      \label{3.1}
d^{D} k \int dz A_{\mu}(k) :e^{ikX(z)} \p _{z} X^{\mu}:
\ee
The regularized ``effective action''
\footnote{We have put quotation marks around the words `effective action'
because it has not really been proved that it is the effective action.
It is plausible, though, because the process of regularization
does subtract the massless poles, at least in the scheme used
in \cite{bspt}.  The regularization used in computing (\ref{3.2})
is the zeta function regularization.}
has been computed exactly \cite{ft1,t}
and is
\be       \label{3.2}
<0\mid 0>_{F} = \sqrt{Det(I+F)}
\ee
The Green's function for the $X$ - fields is known exactly in this limit:
\cite{ac}:
\be        \label{3.3}
\Sigma _{\mu \nu}(z-z^{\prime})=(I-F^{2})^{-1}_{\mu \nu} \ln(z-z^{\prime})
\ee
where $z$ and $z ^{\prime}$ are on the boundary (the real axis).

In the proper time formalism applied to the vector field \cite{bsrg}
the following representation for the vertex operator (\ref{3.1})
was useful in obtaining covariant equations\footnote{This is easily
proved by integrating by parts on $\alpha$ in the second term.  In
\cite{bsfc}, where it was first used, a rather long proof was presented.
That proof involved proving an intermediate result, that is more
general than is required for establishing this result.}
\be      \label{3.4}
A_{\mu}(k)\p _{z} X^{\mu} e^{ikX(z)} =
\int _{0}^{1}d \alpha \p _{z} (A_{\mu} X^{\mu} e^{i\alpha kX})
+i\int _{0}^{1} d \alpha \alpha
k_{[\mu}A_{\nu]}(X^{\mu}\p _{z} X^{\nu} e^{i\alpha kX})
\ee

We can thus calculate the two point function:
\be     \label{3.5}
\int _{0}^{1} d \alpha \alpha \int _{0}^{1} d \beta \beta
<X^{\mu} \p X^{\nu} (z)e^{i\alpha k X} X^{\rho}\p X^{\sigma} (0)
e^{i\beta pX} >_{F} p_{[\rho}A_{\sigma ]}(p) k_{[\mu} A_{\nu ]}(k)
\ee
The coefficient of $A_{\sigma}$ gives the equation of motion.
The subscript ``$F$" indicates that the calculation is done with the
background field $A_{\mu}$ in the action.  We have discarded the
total derivative term, since we know from gauge invariance that
the final result depends only on the field strength.

The result is
\be            \label{3.6}
\sqrt{Det(I+F)}(\Sigma ^{\mu \rho} \p _{z}\p_{z^{\prime}}\Sigma ^{\nu \sigma}
+ \p _{z^{\prime}}\Sigma ^{\mu \sigma} \p _{z}\Sigma ^{\nu \rho})
p_{[\rho}A_{\sigma ]}k_{[\mu}A_{\nu ]}
\ee
Here we have set $k=p=0$ in the exponentials, since we are only interested
in terms that are lowest order in momentum.  Using (\ref{3.3})
one extracts the coefficient of $\frac{\ln z}{z^{2}} A_{\sigma}$
to get
\be   \label{3.7}
\frac{\delta S}{\delta A_{\sigma}} \, = \,
\sqrt{Det(I+F)}(I-F^{2})^{-1\sigma \nu}(I-F^{2})^{-1\lambda \mu}
\p _{\lambda}F_{\nu \mu} = 0  .
\ee
This is the equation of motion obtained by varying $\sqrt{(det(1+F)}$
\cite{ac}.

One can also calculate the Zamolodchikov metric easily
\be   \label{3.8}
<\p_{z} X^{\mu} \p_{z^{\prime}}X^{\nu}(0)>=\frac{1}{z^{2}}
(I-F^{2})^{-1\mu \nu}\sqrt{Det(I+F)} \equiv \frac{G^{\mu \nu}}{z^{2}}
\ee

Comparing this with (\ref{3.7}) one also recovers the $\beta$-function
\cite{ac}:
\be  \label{3.9}
\beta ^{\nu} = (I-F^{2})^{-1\lambda \mu} \p _{\lambda} F_{\nu \mu}
\ee

Thus we conclude from this brief discussion that, as promised, the proper
time equation does give the full equation of motion.  We now turn, in the
next section, to the non trivial situation where the electromagnetic
field is non-uniform.

\newpage
\section{Perturbation Theory for Non Uniform Fields}
\setcounter{equation}{0}

In this section we work out the details of perturbation theory for
evaluating the (gauge invariant) proper time equation.
Thus we have to evaluate
\be      \label{4.1}
\int dk dp \frac{d}{d\ln R}R^{2}<A_{\mu}(k)\p_{z}X^{\mu}e^{ikX(R)}
A_{\nu}(p)\p_{z}X^{\nu}e^{ipX(0)}>
\ee
where the expectation value uses the Polyakov action
perturbed by $\int dz A_{\mu}\p X^{\mu}$. Now, in the original
proper time prescription \cite{bspt} the interactions were
confined between 0 and $z$.  This makes it very similar to
the S-matrix calculation.  However it is possible to extend the
range of integration from $-\infty$ to $+\infty$ using
SL(2,R) invariance.  To see this, consider for e.g. a correlation
involving four vertex operators:
\be    \label{4.2}
<V(z_{1}) V(z_{2}) V(z_{3}) V(z_{4})>
\ee

An SL(2,R) invariant measure is $\int _{-\infty}^{+\infty} d \zi
\int _{-\infty}^{z_{1}} dz_{2} \int _{-\infty}^{z_{2}}dz_{3}
\int _{-\infty}^{z_{3}}dz_{4}$.  Assuming the vertex operators
are on-shell, we can fix one of them to be at R, another at 0,
and a third one at $z$ and the Jacobian is $zR(R-z)$.  A common
choice in S-matrix calculations is $\zi = R \rightarrow \infty
\, , \, z_{2} = 1 $ and $z_{4}=0$. The prescription in the proper time
equation, on the other hand, amounts to the choice $\zi = R, \, \,
z_{2}=z $ and $z_{4} =0$.  Of course as explained in Sec 1, $z$
is also integrated over eventually, in the proper time formalism.
Thus we get
\be     \label{4.3}
\int _{0}^{R} dz \int _{0}^{z}dz_{3}
<V(z_{1}=R) V(z_{2}=z) V(z_{3}) V(z_{4}=0)>
\ee
One could, instead, choose $z_{2} = R, \, \, z_{3} =z $ and $z_{4}=0$
and by SL(2,R) invariance the result would be the same.
In this case $\zi$ would be integrated from $R$ to $\infty$
(because $z_{1} \geq z_{2}$).
Thus, this would contribute a term of the form
\be     \label{4.4}
\int _{R}^{\infty} dz_{1} \int _{0}^{R}dz
<V(z_{1}) V(z_{2}=R) V(z_{3}=z) V(z_{4}=0)>
\ee

Now, in contrast
to an S-matrix calculation, where each vertex operator has associated
with it a definite momentum, in the present case the
vertex operators are of the form
\be     \label{4.5}
V(z) = A_{\mu}(X)\p _{z} X^{\mu} = \int dk A_{\mu}(k) \p _{z} X^{\mu}
e^{ikX}
\ee
The momentum is integrated over and consequently the vertex operators are
identical and indistinguishable.  Thus, by relabelling the
$z_{i}$ 's one sees that the only difference between
(\ref{4.3}) and (\ref{4.4}) is the range of integration of one
of the $z_{i}$ ,i.e.,in (\ref{4.3}) $z_{3}$ is integrated from
0 to $z$, whereas in (\ref{4.4}) $z_{1}$ is integrated from
$R$ to $\infty$.  In both cases $z$ is integrated from 0 to $R$.
We can thus take four terms that are numerically
identical, namely (\ref{4.3}), (\ref{4.4}) and two others with
ranges of integration $(-\infty ,0)$ and $(z,R)$, add them to get
a range of $(-\infty , +\infty )$.  Since we have added four
numerically identical terms we can divide by 4.  In terms of
S-matrix calculation we have added four terms with the
momenta permuted amongst themselves.  Thus we conclude that extending the
range of integration  from $-\infty$ to $+\infty$ has the effect
of doing an S-matrix calculation and symmetrizing on the external
momenta.  Note that the range of integration of the variable
$z$ is always from 0 to $R$.

The above argument depends crucially on SL(2,R) invariance.
Thus we have to assume that the vertex operators
are physical and of dimension one.  For the massless vector this means $k^{2}=
k.A=0$.  The equations we derive are thus {\em a priori} valid only in an
infinitesimal region of momentum space around this.  Any extrapolation
beyond this will require {\em a posteriori} justification based on the final
result.  This is true, of course, for any off-shell extrapolation.

As mentioned in the introduction there are some advantages to thus
extending the range of integartion.  One can, for instance, use
(\ref{1.2}).  This is useful in the zero momentum limit.  In the case
of abelian vectors there is another advantage.  The effective
action is supposed to have the massless poles subtracted out.
  Usually this is taken care of by regularizing the integrals.
In the abelian case, however, since there is no (bare) three-photon
interaction, the vector-vector scattering element does not have
massless poles at all. To explicitly see that the poles are absent,
one has to symmetrize on the external momenta \cite{bsrg}.
This is a little tedious in an actual calculation.  On the other
hand when the range of integration is extended
from $-\infty$ to $+\infty$ one sees this cancellation very easily.
This will be demonstrated explicitly in the next section when we
look at the zero momentum limit.

We now derive the general expression.  To do this we replace
$A_{\mu} \p _{z} X e^{ikX}$ by (\ref{3.4}).  Thus we have to calculate
\be     \label{4.6}
\int _{0}^{1}d\alpha \alpha \int _{0}^{1}d \beta \beta \frac{d}{d\ln R}R^{2}
<k^{[\mu}A(k)^{\nu ]} X^{\mu} \p _{z_{1}}X^{\nu} e^{i\alpha kX(R)}
p^{[\rho}A(p)^{\sigma ]}X^{\rho} \p_{z_{2}} X^{\sigma} e^{i\beta pX(0)}>
\ee
One inserts some number of vertex operators of the form
\be     \label{4.7}
\int _{0}^{1} d \gamma \gamma q^{[\mu} A(q)^{\nu]} X^{\mu}\p_{z}X^{\nu}
e^{i\gamma qX(z)} .
\ee
and integrates all of them from $-\infty$ to $+\infty$, except for one,
which is integrated from $0$ to $R$. If one regulates the
integrals by imposing that the distance of closest approach
between two vertex operators is $a \neq 0$, then one finds that the
on-shell poles are subtracted. \footnote{As mentioned earlier,
in the abelian case this is unnecessary.}

To illustrate all this we turn to the leading correction to Maxwell's
equation, involving two insertions of (\ref{4.7}) in (\ref{4.6}).
Thus consider
\[
\int dk\, dp\, dq\, dl\,
F_{\mu _{1} \nu _{1}}(k) F_{\mu _{2} \nu _{2}}(p) F_{\mu _{3} \nu _{3}}(q)
F_{\mu _{4} \nu _{4}}(l)
\int d \alpha  d\beta d\gamma d\delta \alpha \beta \gamma \delta
\]
\be     \label{4.8}
X^{\mu _{1}} \p _{z_{1}}X^{\nu _{1}} e^{i\alpha kX(z_{1})}
X^{\mu _{2}} \p _{z_{2}}X^{\nu _{2}} e^{i\beta pX(z_{2})}
X^{\mu _{3}} \p _{z_{3}}X^{\nu _{3}} e^{i\gamma qX(z_{3})}
X^{\mu _{4}} \p _{z_{4}}X^{\nu _{4}} e^{i\delta lX(0)}
\ee
There are two types of contractions that one can make in which the
$X^{\mu _{i}}\p X^{\nu _{i}}$ are contracted amongst themselves -
one results in $Tr(F^{4})$ and the other $(TrF^{2})^{2})$. In this
paper we will consider only these .
Let us
consider these two in turn.

{\bf a) $TrF^{4}$:} \\
There are three kinds of contractions:
\[
i) \, \, \frac{1}{z_{1}-z_{2}}\frac{1}{z_{2}-z_{3}}
\frac{1}{z_{3}-z_{4}}\frac{1}{z_{4}-z_{1}}
\]
\[
ii) \, \, \frac{1}{z_{1}-z_{3}}\frac{1}{z_{3}-z_{2}}
\frac{1}{z_{2}-z_{4}}\frac{1}{z_{4}-z_{1}}
\]
\be     \label{4.9}
iii) \, \, \frac{1}{z_{1}-z_{2}}\frac{1}{z_{2}-z_{4}}
\frac{1}{z_{4}-z_{3}}\frac{1}{z_{3}-z_{1}}
\ee

There are $2^{4}$ terms of each type that are obtained by interchanging
$\mu _{i}$ with $\nu _{i}$ in each contraction.  Each of these
are multiplied by the factor $M$ defined below:
\be     \label{4.10}
M = \mid z_{1}-z_{2} \mid ^{\alpha \beta k.p}
    \mid z_{2}-z_{3} \mid ^{\gamma \beta q.p}
    \mid z_{3}-z_{4} \mid ^{\gamma \delta q.l}
\ee
\[
    \mid z_{1}-z_{3} \mid ^{\alpha \gamma k.q}
    \mid z_{2}-z_{4} \mid ^{\delta \beta p.l}
    \mid z_{1}-z_{4} \mid ^{\alpha \delta k.l}
\]
Since $z_{1}$ and $z_{4}$ are fixed there is really nothing to distinguish
between them.  Thus (i) and (ii) (of (\ref{4.9})) are identical. \\
${\bf b) (TrF^{2})^{2}:}$ \\
Again there are three kinds of contractions.
\[
i) \, \, [\frac{1+\ln (z_{1}-z_{2})}{(z_{1}-z_{2})^{2}}]
[\frac{1+\ln (z_{3}-z_{4})}{(z_{3}-z_{4})^{2}}]
\]
\[
ii) \, \, [\frac{1+\ln (z_{1}-z_{3})}{(z_{1}-z_{3})^{2}}]
[\frac{1+\ln (z_{2}-z_{4})}{(z_{2}-z_{4})^{2}}]
\]
\be     \label{4.11}
iii) \, \, [\frac{1+\ln (z_{1}-z_{4})}{(z_{1}-z_{4})^{2}}]
[\frac{1+\ln (z_{2}-z_{3})}{(z_{2}-z_{3})^{2}}]
\ee
Each of these is also multiplied by the factor $M$. Once again (i)
and (ii) give identical results due to the indistinguishability
of $z_{1}$ and $z_{4}$.

We now turn to the actual evaluation of the integrals.  We will only do
a few representative examples.  In doing the integrals it is important
to keep in mind that it is the absolute value $\mid z_{i} - z_{j}\mid$
that occurs in the argument of the logarithms and in $M$, whereas
in the denominator of (\ref{4.9}) it is $(z_{1}-z_{j})$ - {\em without}
the absolute value - which can be positive or negative, depending
on whether $z_{i} \geq $ or $\leq z_{j}$.

Let us consider a typical term in a)(i):
\be     \label{4.12}
\int _{0}^{z_{1}} d z_{2}\int _{-\infty}^{+\infty} dz_{3}
\mid z_{1}-z_{2} \mid ^{ k.p} (z_{1}-z_{2})^{-1}
    \mid z_{2}-z_{3} \mid ^{ q.p}(z_{2}-z_{3})^{-1}
    \mid z_{3}-z_{4} \mid ^{ q.l} (z_{3}-z_{4})^{-1}
\ee
\[
    \mid z_{1}-z_{3} \mid ^{ k.q}
    \mid z_{2}-z_{4} \mid ^{ p.l}
    \mid z_{1}-z_{4} \mid ^{ k.l} (z_{1}-z_{4})^{-1}
\]
We have used $k$ for $\alpha k$, $p$ for $\beta p$ etc for notational
simplicity.  The factors $\alpha , \beta ...$ will have to be restored
at the end.  We have further set $z_{4}=0$.  These integrals
can be expressed in terms of Beta functions and hypergeometric functions.
For simplicity let us choose a region of momentum space where
\be     \label{constr}
p.l=k.q=0
\ee
This is over and above the mass shell constraint
\be     \label{mshell}
k^{2}=p^{2}=q^{2}=l^{2}=0
\ee
The two,(\ref{constr}) and (\ref{mshell}), imply
\be     \label{4.12.5}
k.p=q.l=-p.q=-k.l
\ee
Note, however, that  $k.p,p.q ...$ are not required
to be small.  In this sense we are going beyond the usual
sigma model perturbation theory, where, because a derivative expansion
is being performed, we are forced to have $k \approx 0$,
and not just $k^{2}\approx 0$.

(\ref{4.12}) can be rewritten as:
\be     \label{4.13}
z_{1}^{-2+k.p+p.q+k.l+q.l}
\int _{0}^{1} dz_{2} (1-z_{2})^{-1+k.p} z_{2}^{-1+p.q+q.l}
\int _{-\infty}^{+\infty} dz_{3} (1-z_{3})^{-1+p.q} z_{3}^{-1+q.l}
\ee
where we have performed the usual rescaling of $z_{2}$ and $z_{3}$.
The integrals can be easily done (see Appendix A) and the result is
\be   \label{4.14}
z_{1}^{-2+k.p+p.q+k.l+q.l}
B(k.p,p.q+q.l)
\ee
\[
[-B(1-p.q-q.l,p.q) -B(1-p.q-q.l,q.l)+B(p.q,q.l)]
\]
The logarithmic deviations from the canonical $\frac{1}{z^{2}}$ scaling
is proportional to $(k.p +k.l +p.q +q.l)$ which, by(\ref{4.12.5}), is zero.
However, the first Beta function has a pole $\frac{1}{p.q+q.l}$.
Thus we get $(\frac{k.p+k.l}{p.q+q.l} +1)$.  Since $k.p=q.l$ and
$k.l=p.q$, we actually get a non-zero result, namely 2.
Thus the final answer is
\[
\int dk dp dq dl
F_{\mu _{1} \nu _{1}}(k) F_{\nu _{1} \nu _{2}}(p) F_{\nu _{2} \nu _{3}}(q)
F_{\nu _{3} \mu _{1}}(l)
\]
\be     \label{4.15}
\int d \alpha  d\beta d\gamma d\delta \alpha \beta \gamma \delta
\ee
\[
2[-B(1-p.q-q.l,p.q) -B(1-p.q-q.l,q.l)+B(p.q,q.l)] .
\]

As mentioned earlier, one should really regularize these integrals
and thus subtract any massless poles that might be present in (\ref{4.15}).
However it is easy to see using an expansion of the Beta function (Appendix
B) that the would be poles, in fact, cancel.  The leading term
is $-3(p.q+q.l)\zeta (2)$ (which is actually zero if we use (\ref{4.12.5})).
The final answer (\ref{4.15}) is valid for finite values of momenta
as long as the restriction (\ref{constr}) is satisfied.  Furthermore
it is assumed that (\ref{mshell}) is also approximately satisfied
((\ref{4.12.5}) is then automatic). Note, however, that (\ref{mshell})
was never used in evaluating the integral.  It was only used to
motivate the proper time equation.
In principle, one can adopt the viewpoint that the
equation is correct even when the fields are off-shell.  In that case
the only restriction is (\ref{constr}).

One can proceed to evaluate in a similar manner all the terms in
(\ref{4.9}) and (\ref{4.11}).  We will write down the results for
a couple more of the terms, since they will be used in the next
section for extracting the zero momentum limit.  Let us consider
two more terms in ai)(\ref{4.9}).

\be     \label{4.16}
\frac{1}{(z_{1}-z_{2})^{2}}\ln (z_{2}-z_{3})\frac{1}{(z_{3}-z_{4})}
\frac{1}{(z_{4}-z_{1})}
\ee
\be     \label{4.17}
\frac{1}{(z_{1}-z_{2})^{2}}\ln (z_{4}-z_{1})\frac{1}{(z_{2}-z_{3})}
\frac{1}{(z_{3}-z_{4})}
\ee
It is understood that they are to be multiplied by $M$  (\ref{4.10}).
Consider (\ref{4.16}) first.  We replace the logarithm by
$(z_{2}-z_{3})^{\mu}$ and at the end we will pick the term that is
linear in $\mu$.  The result of doing the integral is ($z_{4}=0$):
\be     \label{4.18}
z_{1}^{-2+k.p+p.q+k.l+q.l +\mu}
\{B(-1+k.p,1+\mu +p.q+q.l)
\ee
\[
[-B(-q.l -p.q-\mu , q.l)+B(1+\mu +p.q, q.l)
+B(-q.l -p.q -\mu , 1+ \mu +p.q)]\}
\]
It is easy to see that, when (\ref{4.12.5}) is satisfied, the coefficient
of $\mu \ln z_{1}$ is the term in curly brackets with $\mu$ set to 0.
Once again it can be checked that the massless poles inside the
square brackets cancel. The appearance of a pole in the first factor
is also deceptive because it cancels against a zero.
Here, and later, we have omitted writing down explicitly
all the accompanying
field strength factors and the
integrals over $\alpha , \beta ...$ etc.

Next, consider (\ref{4.17})
We get $(z_{4}=0)$:

\be     \label{4.19}
z_{1}^{-2}\ln z_{1}
\{B(-1+k.p,p.q+q.l)
\ee
\[
[-B(1-p.q-q.l,p.q) -B(1-p.q-q.l,q.l)+B(p.q,q.l)]\}
\]
The expression in curly brackets is the answer.
Once again regularization is unnecessary - the pole terms cancel.
We will see this explicitly when we study the zero momentum limit
in the next section.

Let us now turn to terms that contribute to $(TrF^{2})^{2}$ given
in (\ref{4.11}).\\
{\bf bi)}
\be     \label{4.20}
 [\frac{1+\ln(z_{1}-z_{2})}{(z_{1}-z_{2})^{2}}]
[\frac{1+\ln (z_{3}-z_{4})}{(z_{3}-z_{4})^{2}}] \times M
\ee
We consider the integral in the same approximation (\ref{constr}).
\be     \label{4.21}
\int _{0}^{z_{1}} dz_{2}\int _{-\infty}^{+\infty} dz_{3}
(z_{1}-z_{2})^{-2+k.p+\mu}  (z_{2}-z_{3})^{p.q} (z_{3})^{-2+q.l+\nu}
(z_{1})^{k.l}
\ee
The answer is given by the coefficients of
1,$\mu , \nu $ and $ \mu \nu$.
The result for the coefficient of $\ln z_{1}$ is:
\[
(\mu +\nu)B(-1+k.p+\mu ,p.q+q.l+\nu )
\]
\[
[B(1-p.q-q.l-\nu , -1 +q.l+\nu )
+B(1-p.q-q.l - \nu , 1+p.q)
\]
\be     \label{4.22}
+B(-1+q.l+\nu , 1+p.q )]
\ee
One has to expand the Beta function in powers of $\mu , \nu$ to
get the final answer.

{\bf biii)}
\be     \label{4.23}
 [\frac{1+\ln (z_{1}-z_{4})}{(z_{1}-z_{4})^{2}}]
[\frac{1+\ln (z_{2}-z_{3})}{(z_{2}-z_{3})^{2}}] \times M
\ee
Using a different approximation
\be     \label{4.24}
p.l=k.q=-q.l=-k.p
\ee
we get
\be     \label{4.25}
\int _{0}^{z_{1}} dz_{2}\int _{-\infty}^{+\infty} dz_{3}
(z_{1}-z_{2})^{k.p}  (z_{2}-z_{3})^{-2+p.q+\mu } (z_{3})^{q.l}
(z_{1})^{k.l}
\ee
We have introduced, as before, $\mu$ for the logarithm.
The result is
\be     \label{4.26}
z_{1}^{-2+k.p+p.q+k.l+q.l+ \mu}(1+\ln z_{1})
\{B(1+k.p,p.q+q.l+\mu )
\ee
\[
[B(1-p.q-q.l-\mu , 1+q.l)+B(1-p.q-q.l-\mu ,
-1+p.q +\mu ) +B(1+q.l , -1+p.q +\mu )]\}
\]
One has to extract the coefficient of $\ln z_{1}$ from the above,
keeping terms independent of $\mu$, and also linear in $\mu$.  We will not
do this here since it is tedious and unilluminating. The low energy
limit is worked out in the next section.

Thus the main results of this section are, (\ref{4.15}),(\ref{4.18}),
(\ref{4.19}),(\ref{4.22}) and (\ref{4.26}), which are some
representative coefficients of the form $TrF^{4}$ and $(TrF^{2})^{2}$
to Maxwell's action.  In the region of momentum space satisfying
(\ref{constr}), these are exact to all orders in derivatives
and are therefore valid for finite values of momenta.  To complete the
calculation one has to pick out linear and bilinear terms in $\mu ,\nu$.
This can easily be done by expanding the Beta function in a power series.
Alternatively they can be expressed in terms of $\Psi$-functions
defined as \cite{gr}
\be     \label{4.27}
\Psi (x)= \frac{d}{dx} \ln \Gamma (x)
\ee
Thus, for instance, to pick the piece linear in $\mu$ in
$B(x+\mu , y+\nu )$ one can use:
\be     \label{4.28}
\left.\frac{dB}{d \mu} \right| _{\mu =0} = B(x,y)[\Psi (x) - \Psi (x+y)]
\ee
Finally one has to perform the $\alpha , \beta , \gamma ,\delta $
integrals - these are the parameters used in (\ref{3.4}).  In calculating
correlation functions of the type in (\ref{4.8}), after Wick
contraction of
the $X^{\mu _{i}} \p _{i} X^{\nu _{i}}$ amongst themselves, one
is left with a vacuum expectation value of four exponentials.
Momentum conservation gives a constraint
\be     \label{4.29}
\alpha k^{\mu}+ \beta p^{\mu}+\gamma q^{\mu}+\delta l^{\mu} =0
\ee
Thus, (\ref{4.12.5}) holds even when the parameters
$\alpha , \beta ...$ are
included :
\[
\alpha k. \beta p = \gamma q. \delta l
\]
We can therefore replace $\delta l ^{\mu}$ everywhere by
$-(\alpha k ^{\mu} + \beta p ^{\mu} + \gamma q^{\mu})$.  The integral
$\int d^{D}l$ can be replaced by $\int d^{D}(\delta l)$ (which is 1 because
of the momentum conservation delta function) multiplied by $\delta ^{-D}$.
The integral over the parameter $\delta$ is then an overall constant
(infinite) factor that is common to every term and can be
set to 1.

To summarize this section, we have described a well defined prescription
for a (covariant) perturbative evaluation of the proper time equation.

\newpage
\section{Zero Momentum Limit}
\setcounter{equation}{0}
The purpose of this section is to consider the zero momentum
limit of the results of the previous section and compare it with
known results\cite{ac,ft1,t}.  This will provide a non trivial
check on the details of the perturbation scheme presented in
the last section.  It also serves to bring out certain subtleties
in the process of taking the zero momentum limits in these
kinds of calculations that can be a bit perplexing.

If  the zero momentum limit is taken directly we run into
problems.  Taking the zero momentum limit is equivalent to
setting the factor $M=1$ (\ref{4.10}).  If we consider the original
proper time prescription of having a range of integration for
$z_{i}$ between $z_{i-1}$ and $z_{i+1}$ we run into integrals of the type
\be     \label{5.1}
\int _{u}^{z}dw \frac{1}{z-w}\frac{1}{w-u}
\ee
This is of course divergent at both ends, which is symptomatic of the
infrared divergence due to massless poles $\frac{1}{p^{2}}$,
as $p\rightarrow 0$.  In the S-matrix calculation
(in the abelian case),
these poles cancel when
we symmetrize on the external momenta.  We have already seen that this is
equivalent to extending the range of integration
from $-\infty $ to $+\infty$.
Thus we get

\be     \label{5.2}
\int _{-\infty} ^{+\infty} dw \frac{1}{z-w} \frac{1}{w-u}
\ee
If we introduce infrared and ultraviolet regulators, we end
up with expressions of the type $\ln (R/a) - \ln (R/a) =0$!
Actually the cure is obvious, because what we really have
is $\ln (1) = 0,2\pi i,...$Thus we need to be very careful about the
$i\epsilon$ prescription in (\ref{5.2}).  In fact we should use
the principal value prescription (this can be derived from the
momentum representation, see Appendix C), and we get
\be     \label{5.3}
1/4 \int _{-\infty} ^{+\infty} dw
[\frac{1}{z-w-i\epsilon }+\frac{1}{z-w+i\epsilon}][\frac{1}{w-u-i\epsilon}
+\frac{1}{w-u+i\epsilon}]  = -\pi ^{2} \delta (z-u)
\ee
This agrees with (\ref{1.2}).  The moral of this simple calculation
is that one must be careful in regulating expressions like
(\ref{5.1}) or (\ref{5.2}).  One way of regulating (\ref{5.2}) is to
introduce $i \epsilon$ as above.  Another way is to restore the factor
$M$ of (\ref{4.10}) and take the limit $p^{\mu} \rightarrow 0$ at the end.
The $\beta$-function calculation \cite{ac} uses (\ref{5.3}).  This works
well for the $TrF^{4}$ term but not for the $(TrF^{2})^{2}$ term as
we see below.
Thus, consider the terms of (\ref{4.9}).  We have the result (\ref{4.15})
for (ai).  Expanding the Beta functions in powers of momenta one
finds for the expression in square brackets (see Appendix B):
\be     \label{5.4}
[B(p.q,q.l)-B(1-p.q-q.l,p.q)-B(1-p.q-q.l,q.l)] \, = \, -3(p.q+q.l)\zeta (2)
\ee
The pole terms have cancelled.  (\ref{5.4}) vanishes if we use (\ref{4.12.5}),
but even otherwise it vanishes in the zero momentum limit.

Similarly (\ref{4.18}) gives for the expression in square brackets
$-3(\mu + p.q)\zeta (2)$, and expanding the prefactor also, we get
for the coefficient of $\mu$,
\be     \label{5.5}
\frac{1}{k.p-1} \frac{1}{k.p}(k.p+p.q+q.l )
[1-k.p(p.q+q.l)\zeta (2)][-3( p.q) \zeta (2)]
\ee
Clearly this also vanishes in the zero momentum limit.

Finally, we turn to (\ref{4.19}).  We get:
\be     \label{5.6}
[\frac{p.q+q.l+k.p -1}{k.p-1}][\frac{1}{p.q+q.l} + \frac{1}{k.p}]
(-3)(p.q+q.l)\zeta (2)
\ee
Although $p.q+q.l \approx 0$, the pole cancels against the zero and we get
\be     \label{5.7}
-3\zeta (2) =- \frac{\pi ^{2}}{2}
\ee
Thus of the three terms considered (all from (ai)), only the last
one, (\ref{4.17}), containing a $\ln \mid z_{1}-z_{4}\mid$ factor,
contributes to the zero momentum limit.

If we use (\ref{5.3}) this is obvious:
\[
\int _{0}^{z_{1}}dz_{2} \int _{-\infty}^{\infty}dz_{3}
\frac{1}{z_{1}-z_{2}}\frac{1}{z_{2}-z_{3}}
\frac{1}{z_{3}-z_{4}}\frac{1}{z_{4}-z_{1}}
\]
\[
= -\pi ^{2} \int _{0}^{z_{1}}dz_{2} \frac{1}{z_{1}-z_{2}}
\frac{1}{z_{4}-z_{1}}
  \delta (z_{2})
\]
\be     \label{5.8}
=\frac{\pi ^{2}}{2} \frac{1}{z_{1}^{2}}
\ee
The delta function contributes only 1/2, because we only integrate
between 0 and $z_{1}$.  Evidently this does not have a $\ln z_{1}$
dependence, hence it does not contribute to the proper time equation.

(\ref{4.17}) is

\[
\int _{0}^{z_{1}}dz_{2} \int _{-\infty}^{\infty}dz_{3}
\frac{1}{(z_{1}-z_{2})^{2}}\frac{1}{z_{2}-z_{3}}\frac{1}{z_{3}-z_{4}}
\ln \mid z_{4} - z_{1}\mid
\]
\[
=- \pi ^{2} \int _{0}^{z_{1}}dz_{2}\frac{1}{(z_{1}-z_{2})^{2}}
  \delta (z_{2}) \ln  z_{1}
\]
\be     \label{5.9}
=\, \frac{-\pi ^{2}}{2} \frac{1}{z_{1}^{2}} \ln  z_{1}
\ee
This contributes $-\frac{\pi ^{2}}{2}$ to the proper time
equation, in agreement with (\ref{5.7}).  Thus of the
$2^{4}=16$ terms in(\ref{4.9})(ai), only four contribute
to the proper time equation.  The same is obviously true for
(aii).  Finally it is easy to see that (aiii) does not contribute at all.
Thus we get a total of (with another minus sign from the Lorentz
index contraction):
\be   \label{5.9.5}
4\pi ^{2} TrF^{4}
\ee

We now turn to (b) (\ref{4.11}).  Consider bi).  The exact result
is given in (\ref{4.22}).  The expression in square brackets
gives
\be     \label{5.10}
3p.q(\nu + q.l+p.q)\zeta (2)
\ee
and thus the full result is
\be     \label{5.11}
\ln  z_{1} (\mu + \nu ) [ \frac{1}{p.q+q.l+\nu } + \frac{1}{k.p+\mu}]
3p.q(\nu +q.l +p.q) \zeta (2)
\ee
Clearly we have a momentum independent piece:
\be     \label{5.12}
3\mu \nu \frac{p.q}{k.p} \zeta (2) = -3 \zeta (2) = -\frac{\pi ^{2}}{2}
\ee
Similarly bii) gives $-\frac{\pi ^{2}}{2}$.

As for biii), we have
(\ref{4.26}) which simplifies to
\be   \label{5.13}
\frac{1}{p.q+q.l+\mu }3q.l(p.q+q.l+\mu ) \zeta (2)
\approx 3q.l\zeta (2)
= \frac{\pi ^{2}}{2} q.l
\ee
which vanishes in the zero momentum limit.

Thus as far as $(TrF^{2})^{2}$ is concerned we get a total of
\be     \label{5.13.5}
-\pi ^{2} (Tr F^{2})^{2}
\ee
On the other hand, if we try to do the same calculation using
(\ref{5.3}) we get for bi)
\be     \label{5.14}
\int _{0} ^{z_{1}} \frac{\ln (z_{1}-z_{2})}{(z_{1}-z_{2})^{2}}
\int _{-\infty} ^{\infty}\frac{\ln (z_{3}-z_{4})}{(z_{3}-z_{4})^{2}}
\ee
We can obviously integrate by parts on $z_{3}$ and using (\ref{1.2}), get
$\delta(z_{4}-z_{4}) = \delta (0) $ - which gives a divergent answer.
Actually, if we use an infrared regulator in the propagator, \footnote
{Use $\frac{1}{p^{2}+m^{2}}$ instead of $\frac{1}{p^{2}}$}  it
is easy to see that the delta function acts only on non-constant
functions and is zero otherwise.  We therefore get zero or infinity
depending on the regularization.  Thus this method, which worked very well
for the $TrF^{4}$ term, gives ambiguous (wrong) answers here.

The final result, for the zero momentum limit is (combining (\ref{5.9.5})
and(\ref{5.13.5})):
\be     \label{5.15}
4\pi ^{2}[TrF^{4} -\frac{1}{4} (TrF^{2})^{2}]
\ee
To be more precise, it is the coefficient of $A^{\mu}(k)$ in the above
expression.  This clearly agrees with the Born-Infeld results
(\ref{1.5}) \cite{ac,ft1,t}.
\newpage
\section{Conclusions}
\setcounter{equation}{0}
In this paper we have discussed, in some detail, the proper time
equation for the electromagnetic field in the open string.
In Section 2 we have tried to give an overview
of the proper time equation and its intimate connection with the
Zamolodchikov metric.  In Section 3 we illustrated this with the
exactly calculable case of a uniform electromagnetic field.
We showed that the proper time equation gives the full (Born-Infeld)
equation of motion.  In Section 4 we gave a systematic prescription
for evaluating the covariant proper time equation in the general
momentum dependent case.  We illustrated it by calculating some of the
leading corrections to Maxwell's equation.  We showed that one can write
down in some cases, closed form expressions, that are exact in their
momentum dependence for {\em finite} values of momenta, rather than
for infinitesimal values as in the sigma model case.  Furthermore
one sees the presence of massive poles.  Thus the radius of
convergence of the momentum expansion is manifest.  Finally in Section 5
we discussed the zero momentum limit and showed that it agrees with the
Born-Infeld results.  This provides a non trivial check on the method.

We thus conclude that the proper time equation can, using the
perturbation scheme described in this paper, be used for a
systematic evaluation of the equation of motion.  The equations
are covariant.  It has also been applied to the non-abelian
case \cite{bsfc,bsrg}.  We believe it can be applied to the
massive modes also, but the issue of gauge invariance at the
interacting level has not been addressed.

  The connection with the Zamolodchikov metric is interesting.
The geometric significance of this metric has not been really
explored.  Since the metric is not a coordinate invariant
object, there must be a better, coordinate invariant
version of the proper time equation.  The `coordinates',
here, are the D-dimensional fields, not $X^{\mu}$.
Perhaps some of the results of \cite{hs,rsz} can be fruitfully
applied here to unearth new results in string theory.

  The background independent formalism of \cite{w,wl} uses
something very similar to the Zamolodchikov metric.  The equation
of motion also has a strong resemblence to the proper time equation.  The
main difference is that BRST transformations are being used rather
than renormalization group transformations. It will be interesting
to understand the precise connection.

Finally, one hopes that some generalization of the techniques
presented here will be applicable to the massive modes and will shed
some light on the underlying principles of string theory.

\newpage
\appendix{\Large \bf APPENDIX}
\section{Evaluation of Integrals}
The first integral in (\ref{4.13}), over $z_{2}$ is just the first
Beta function of (\ref{4.14}).  The second one can be written as
\be
\underbrace{
\int _{-\infty}^{0} dz_{3} (1-z_{3})^{-1+p.q}z_{3}^{-1}\mid z_{3}\mid ^{q.l}
}_{(a)}
\ee
\[
+
\underbrace{
\int _{0}^{1} dz_{3} (1-z_{3})^{-1+p.q}z_{3}^{-1+q.l}
}_{(b)}
\]
\[
+ \underbrace{
\int _{1}^{\infty} dz_{3} (1-z_{3})^{-1}\mid 1-z_{3}\mid ^{p.q}z_{3}^{-1+q.l}
}_{(c)}
\]
Note that one has to be careful about whether to use the absolute value
or not.
\be
(a) \, = \, - \int _{-\infty}^{0} dz_{3}
(1-z_{3})^{-1+p.q}\mid z_{3}\mid ^{-1+q.l}
 = \, -\int _{0}^{\infty} d\mid z_{3}\mid
(1+\mid z_{3}\mid )^{-1+p.q}\mid z_{3}\mid ^{-1+q.l}
\ee
Using \cite{gr}
\be
\int _{u}^{\infty} (x+\beta )^{-\nu} (x-u)^{-1+\mu } dx =
(u+ \beta ) ^{\mu - \nu}B(\nu - \mu , \mu )
\ee
we get
\be
(a) \, = \, -B(1-p.q-q.l,p.q)
\ee
The remaining integrals can be done similarly to get the result
(\ref{4.14}).
\section{Expansion of Beta Functions}
\setcounter{equation}{0}
The following expansion is useful:
\be
 \frac{\Gamma (1+\mu ) \Gamma (1+ \nu )}{\Gamma (1+\mu + \nu )}
 \, = \, 1 - \mu \nu \zeta (2) + higher \, order
\ee
Using this and the recursion relation for the Gamma function, one finds:
\be
B(p.q,q.l) = \frac{1}{p.q} + \frac{1}{q.l} - (p.q +q.l) \zeta (2)
\ee
\be
-B(1-p.q-q.l,q.l) = - \frac{1}{q.l} - (p.q +q.l) \zeta (2)
\ee
\be
-B(1-p.q-q.l,p.q) = - \frac{1}{p.q} - (p.q +q.l) \zeta (2)
\ee
Adding, we see that the poles cancel and we get
\be
-3(p.q+q.l)\zeta (2)
\ee
\section{i$\epsilon$ Prescription for Greens Function}
\setcounter{equation}{0}
The momentum representation (with some normalization)for the
Greens function in two dimensions is:
\be
G(\tau , \tau ^{\prime} ) \, = \, 1/2 \int _{-\infty}^{+\infty}dp_{0}
\frac{1}{\sqrt{p_{0}^{2} +m^{2}}}e^{ip_{0}(\tau - \tau ^{\prime})}
\ee
Here $m$ is an infrared regulator.
Thus
\be     \label{c2}
\frac{\p}{\p \tau}G(\tau , \tau ^{\prime} ) \, =
\, 1/2 \int _{-\infty}^{+\infty}dp_{0}
ip_{0}\frac{1}{\sqrt{p_{0}^{2} +m^{2}}}e^{ip_{0}(\tau - \tau ^{\prime})}
\ee
In the limit $m \rightarrow 0$ this becomes:
\be
= \, i/2 \int _{0}^{\infty} dp_{0} e^{ip_{0}[(\tau - \tau ^{\prime})+i
\epsilon ]} \, +
 \, i/2 \int _{-\infty}^{0} dp_{0}(-1) e^{ip_{0}[(\tau - \tau ^{\prime})-i
\epsilon ]}
\ee
where we have added the $i \epsilon$ to ensure convrgence.
Thus we get
\be
\frac{-i}{2}[\frac{1}{\tau -\tau^{\prime}+i\epsilon}
+ \frac{1}{\tau -\tau ^{\prime}-i\epsilon}] \, ,
\ee
which is the Principal Value prescription.
Note also from (\ref{c2}) that as $p \rightarrow 0$ ,
$\p _{\tau} G(\tau , \tau ^{\prime}) \rightarrow 0$ as long as $m \neq 0$.

\end{document}